\documentclass[review]{elsarticle}
\usepackage[utf8]{inputenc}

\usepackage{hyperref}

\usepackage{tikz}
\usepackage{amsmath}

\journal{Nuclear Instruments and Methods in Physics Research A}

\begin{document}

\begin{frontmatter}

\title{Generative Surrogates for Fast Simulation: TPC Case}



\author[HSE,YSDA]{Fedor Ratnikov\corref{mycorrespondingauthor}}
\cortext[mycorrespondingauthor]{Corresponding author}
\ead{fedor.ratnikov@gmail.com}

\author[HSE]{Artem Maevskiy}
\ead{artem.maevskiy@cern.ch}

\author[JINR]{Alexander Zinchenko}
\ead{alexander.zinchenko@jinr.ru}

\author[PNPI]{Victor Riabov}
\ead{riabovvg@gmail.com}

\author[HSE]{Alexey Sukhorosov}
\ead{assukhorosov@edu.hse.ru}

\author[HSE]{Dmitrii Evdokimov}
\ead{dmevdok@gmail.com}

\address[HSE]{HSE University, 20 Myasnitskaya Ulitsa, Moscow, Russia}
\address[YSDA]{Yandex School of Data Analysis, 11-2 Timura Frunze Street, Moscow, Russia}
\address[JINR]{Joint Institute for Nuclear Research, 6 Joliot-Curie St, Dubna, Moscow Oblast, Russia}
\address[PNPI]{Petersburg Nuclear Physics Institute, 1, mkr. Orlova roshcha, Gatchina, Leningradskaya Oblast, Russia}

\begin{abstract}
Simulation of High Energy Physics experiments is widely used, necessary for both detector and physics studies. Detailed Monte-Carlo simulation algorithms are often limited due to the computational complexity of such methods, and therefore faster approaches are desired. Generative Adversarial Networks (GANs) are well suited for aggregating a number of detailed simulation steps into a surrogate probability density estimator readily available for fast sampling. In this work, we demonstrate the power of the GAN-based fast simulation model on the use case of simulating the response for the Time Projection Chamber (TPC) in the MPD experiment at the NICA accelerator complex. We show that our model can generate high-fidelity TPC responses, while accelerating the TPC simulation by at least an order of magnitude. We describe alternative representation approaches for this problem and also outline the roadmap for the deployment of our method into the software stack of the experiment.
\end{abstract}


\end{frontmatter}


\section{Introduction}
Efficient detector simulation algorithms are crucial for successful planning, optimization and operation of High Energy Physics experiments~\cite{HEPSoftwareFoundation:2017ggl}. The number of samples that can be produced with the existing detailed Monte-Carlo simulation approaches is often limited by the computational complexity of such methods. Faster simulation can be achieved by aggregating a number of detailed simulation steps into a single surrogate generative model that learns the underlying probability density and allows for prompt sampling. Generative Adversarial Networks (GANs)~\cite{Goodfellow:2014upx} is a deep learning method that is well-suited for this goal~\cite{deOliveira:2017pjk}. In our previous work~\cite{Maevskiy:2020ank}, we introduced a GAN-based fast simulation model for the Time Projection Chamber (TPC) in the MPD experiment at the NICA accelerator complex. Here we briefly review our fast simulation approach along with its validation and then describe an alternative representation approach for this problem. We also outline the deployment of our method into the software stack of the experiment.

\section{Fast simulation approach}

Our fast simulation approach for the TPC detector 
consists of building
a GAN that generates the raw detector responses conditioned on the ground truth event content~\cite{Maevskiy:2020ank}. For each heavy- ion collision, MPD TPC response contains approximately $3\cdot10^7$ numbers. To simplify the problem of generating such a large amount of numbers, we factorize the response into an additive decomposition of contributions from small track segments. Within our approach, each track segment contributes to a geometrically localized window, making the target response just an 8-by-16 matrix.

\begin{figure}
    \centering
    \vspace{-0.5cm}
    \begin{tikzpicture}
        \draw (0, 0) node[inner sep=0] {\includegraphics[width=0.41\linewidth]{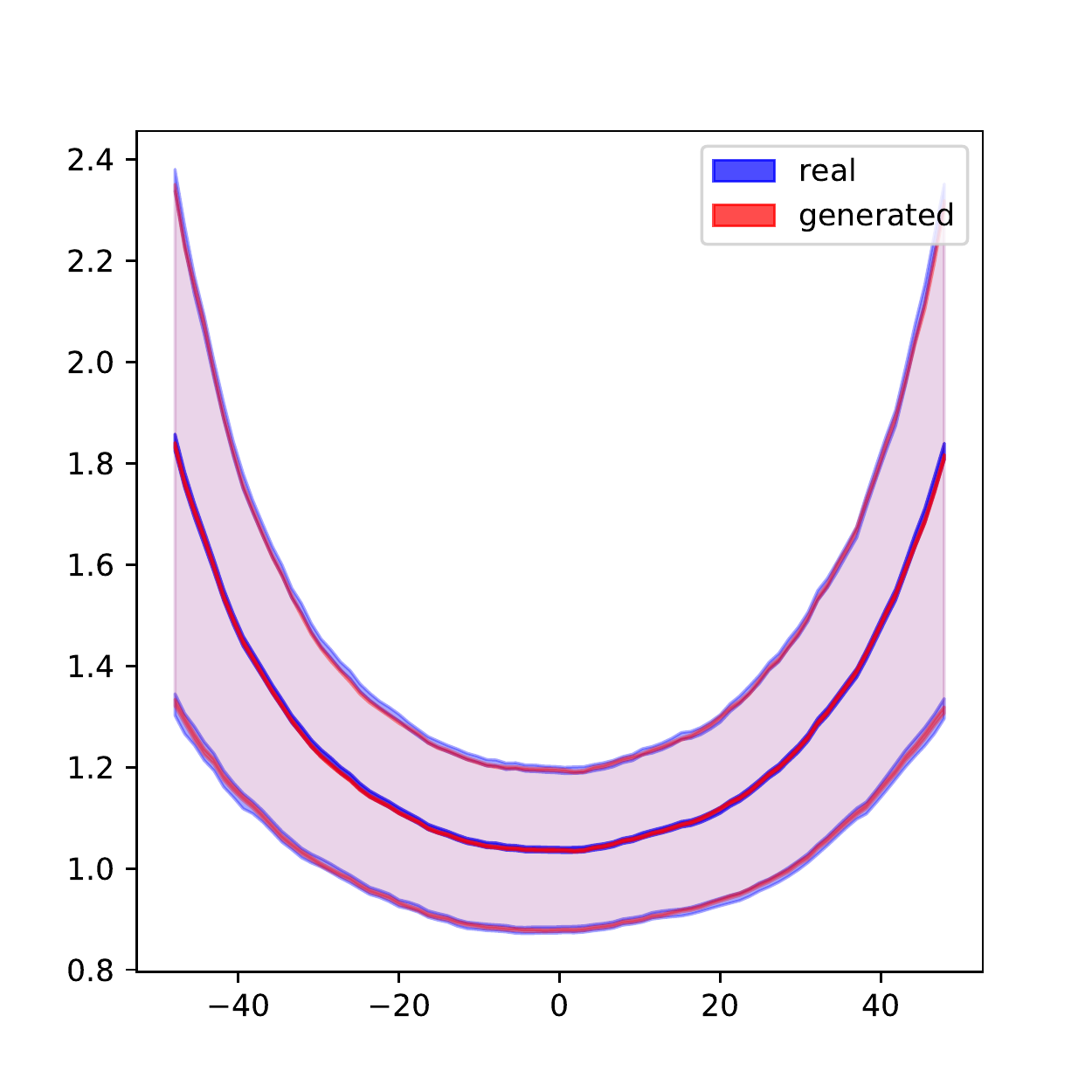}};
        \draw (-1.85, 0) node[rotate=90] {\scriptsize Pad barycenter};
        \draw (0, -1.85) node {\scriptsize Pad coordinate fraction};
    \end{tikzpicture}
    \includegraphics[width=0.525\linewidth,trim=1.25cm -0.5cm 1cm 1cm]{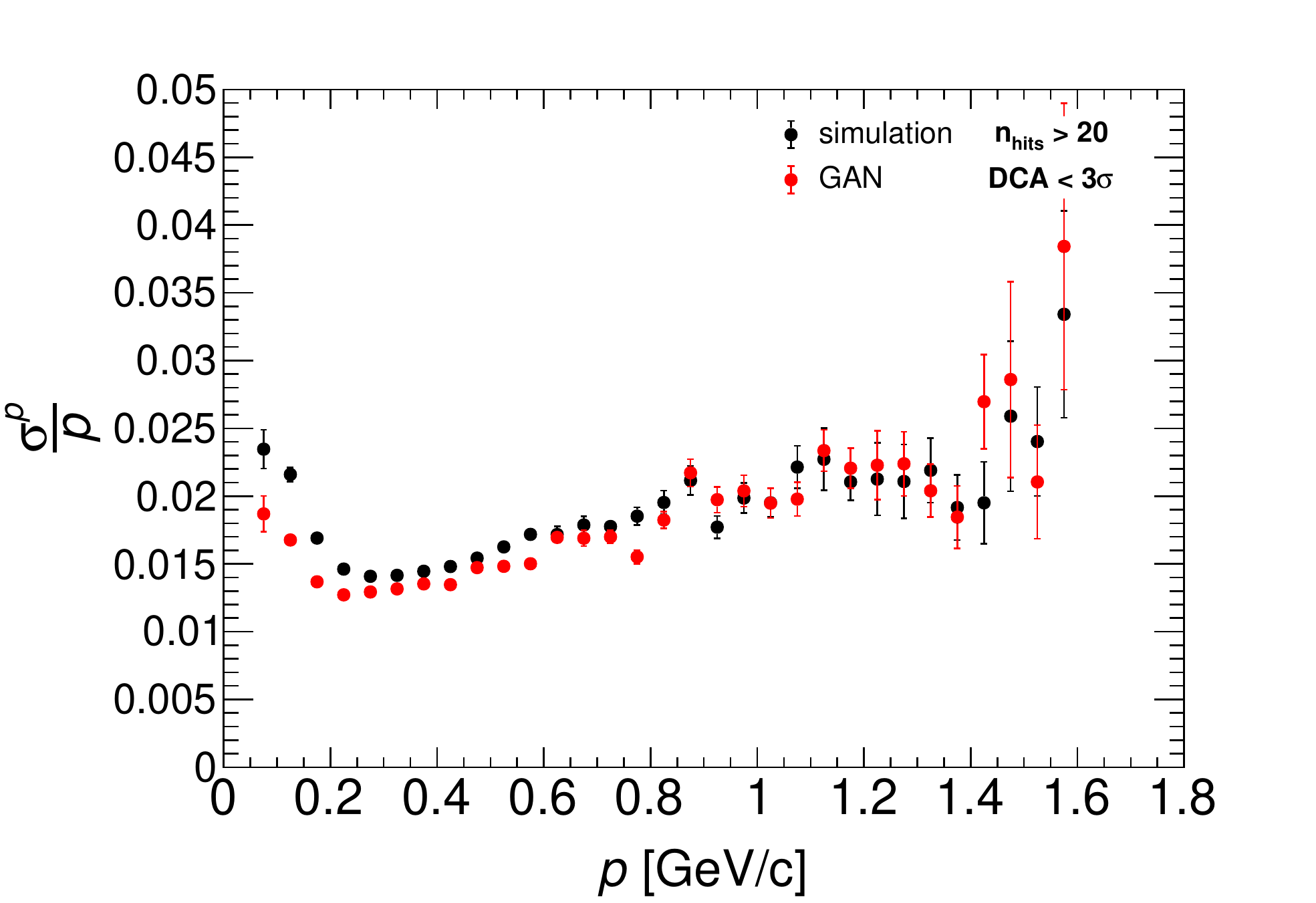}
    \vspace{-0.25cm}
    \caption{(left) Low-level validation profiles. Thick lines correspond to (top to bottom): $\mu+\sigma$, $\mu$, $\mu-\sigma$. Line thickness denotes the statistical uncertainty. (right) Relative momentum resolution as a function of the full momentum.}
    \label{fig:val}
\end{figure}

The built model is validated both at the low level with the raw detector responses and at the high level using reconstructed track characteristics. The low-level validation is done by calculating physically motivated projections from the 8-by-16 response matrix: two barycenters (which are the proxies for the reconstructed cluster coordinates), and the two widths and corresponding covariance (which describe the shape of the signal and therefore define the two-track resolution of the detector). The profiles for the distributions of these quantities as a function of the conditional variables are then built and compared between the model prediction and the detailed simulation, as shown in Fig.~\ref{fig:val}~(left). According to these metrics, our model reproduces the detailed simulation with high accuracy.

The high-level validation is done by integrating our model into the MPD codebase and running the full reconstruction on the GAN-generated responses. The reconstructed track characteristics are then compared with the detailed simulation predictions. An example of such characteristic is shown in Fig.~\ref{fig:val}~(right). Most of the characteristics are reproduced with high fidelity. Some minor discrepancies may be seen, which are well understood and expected to be corrected for in the next iterations of the model development (see~\cite{Maevskiy:2020ank} for more details). The TPC simulation with our model runs $\sim12$ times faster compared to the detailed simulation on the target hardware.

\section{Alternative low-dimensional representation}

Since the most important signal properties are the above-mentioned barycenters and widths, one may train a GAN to generate these low-level characteristics directly and then analytically and deterministically construct the desired pad response matrices from them. In this way, we drastically reduce the dimensionality of the problem and therefore may expect more stable and efficient training, with quality at least as good as in our main approach.

\begin{figure}
    \centering
    \vspace{-0.25cm}
    \begin{tikzpicture}
        \draw (0, 0) node[inner sep=0] {\includegraphics[width=\linewidth]{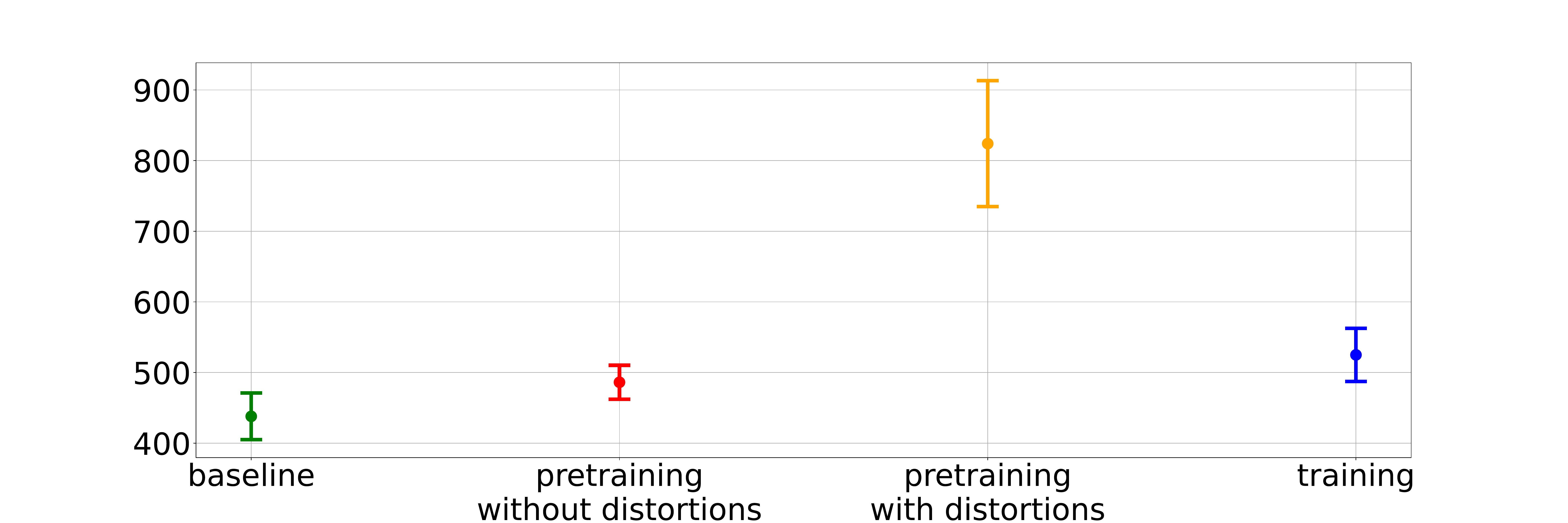}};
        \draw (-4, 0) node[rotate=90] {\scriptsize $\widetilde{\chi^2}$ (lower better)};
    \end{tikzpicture}

    \vspace{-0.25cm}
    \caption{Quality comparison between our main model (the baseline) and the alternative model working in the low-dimensional representation.}
    \label{fig:altmodel}
\end{figure}

We have examined such low-dimensional approach with a choice of the discretized 2D Gaussian function for constructing the response matrices. Unfortunately, the discretization distorts the low-level characteristics, so, in order to account for these distortions, we train the model in two stages. At the first stage, which we call \emph{pretraining}, the model is trained to reproduce the low-level characteristics directly. Then, at the second stage, which we call \emph{training}, we introduce the distorting function in between the generator and discriminator such that the generator learns to account for the distortions automatically. The quality of our best such model, compared with our main approach (the \emph{baseline}), is shown in Fig.~\ref{fig:altmodel}. Contrary to our expectations, we find that this alternative model is less stable in training and requires finer hyper-parameter tuning to achieve a reasonable quality level.

\section{Model deployment pipeline}

\begin{figure}
    \centering
    \includegraphics[width=\linewidth]{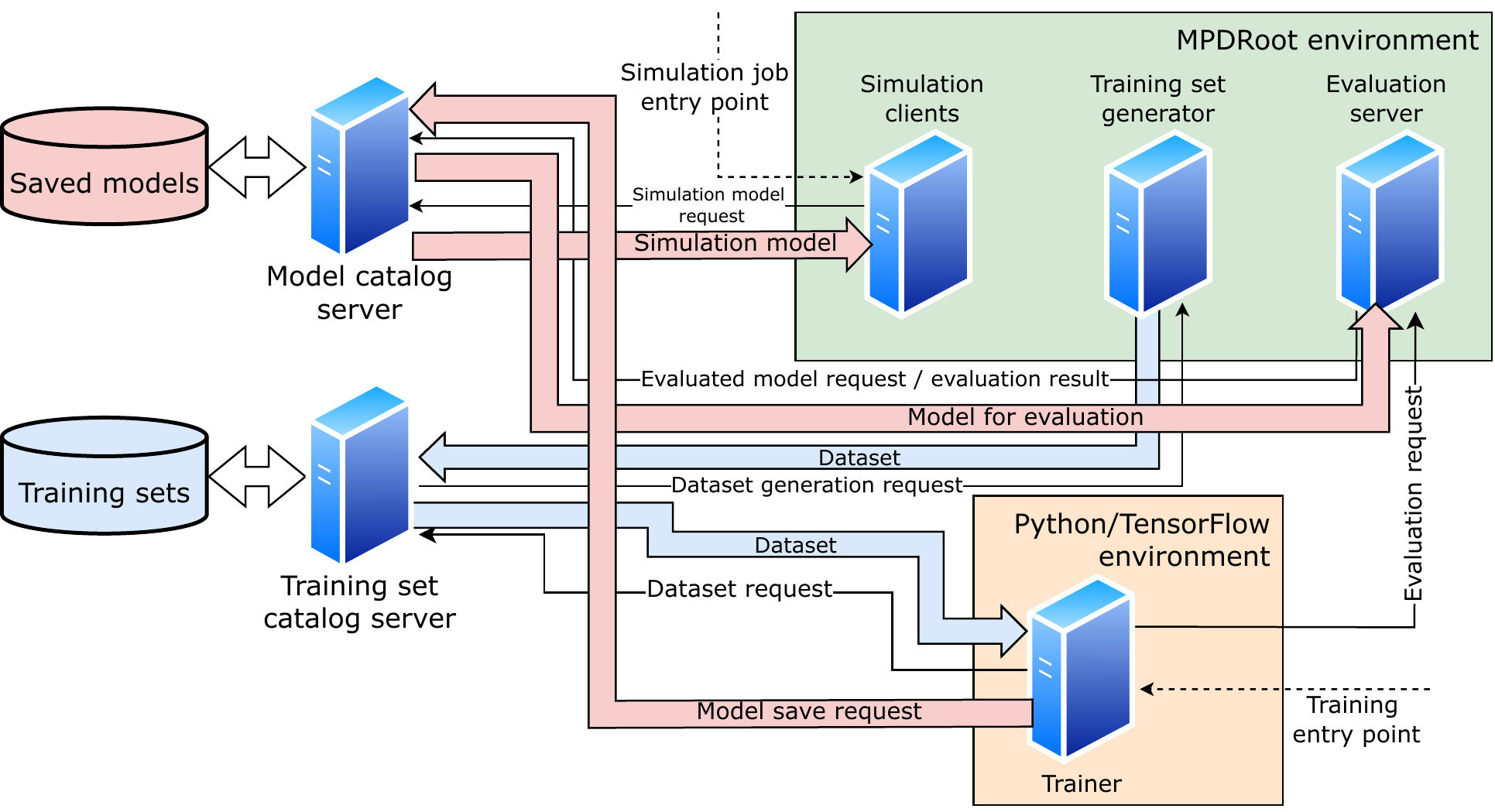}
    \vspace{-0.65cm}
    \caption{The model deployment pipeline.}
    \label{fig:production}
\end{figure}

The simulation configurations are expected to vary as MPD development progresses and running conditions are adapted for. Each new configuration requires our fast simulation model to be re-trained. It is therefore desired to automate the training process as much as possible, including the following steps:
training data generation, model training, model evaluation and selection. Since a multitude of models arise for different configurations, a
model library and database for storage and prompt model retrieval are required. A preliminary scheme for such a pipeline is shown in Fig.~\ref{fig:production}.

A prototype of this scheme has been implemented with the following framework choices. We store the models in the ONNX\footnote{\url{https://onnxruntime.ai/}} format and use ONNXruntime for inference. Airflow\footnote{\url{https://airflow.apache.org/}} is used to manage the workflow with the calculation pipelines described as Directed Acyclic Graphs (DAGs). We use the “Model Registry” component from MLflow\footnote{\url{https://mlflow.org/}} as the model library, which allows us to use REST API to download a model for inference directly in the MPD code.

\section{Summary}

In this work, we have demonstrated our GAN-based fast simulation model for the Time Projection Chamber (TPC) in the MPD experiment at the NICA accelerator complex. Our model can generate high-fidelity TPC responses with an order of magnitude acceleration. We have also described an alternative low-dimensional representation approach for this problem which, contrarily to our expectations, requires more thorough hyper-parameter tuning to achieve a reasonable quality level. Finally, we have shown the strategy for the deployment of our method into the software stack of the experiment.

\bibliography{mybibfile}

\end{document}